\documentclass[aps,prl,twocolumn,superscriptaddress]{revtex4-1}
\usepackage{graphicx}
\usepackage{epstopdf}
\usepackage{color}
\usepackage{bm}

\begin{document}
\title{Projective symmetry of partons in Kitaev's honeycomb model}
\author{Paula Mellado}
\affiliation{School of Engineering and Applied Sciences, 
	Adolfo Ib{\'a}{\~n}ez University,
	Santiago, Chile
}
\author{Olga Petrova}
\affiliation{Max Planck Institute for the Physics of Complex Systems, 01187 Dresden, Germany
}
\author{Oleg Tchernyshyov}
\affiliation{Department of Physics and Astronomy, 
	Johns Hopkins University,
	Baltimore, Maryland 21218, USA
}

\begin{abstract} 
Low-energy states of quantum spin liquids are thought to involve partons living in a gauge-field background. We study the spectrum of Majorana fermions of Kitaev's honeycomb model on spherical clusters. The gauge field endows the partons with half-integer orbital angular momenta. As a consequence, the multiplicities reflect not the point-group symmetries of the cluster, but rather its projective symmetries, operations combining physical and gauge transformations. The projective symmetry group of the ground state is the double cover of the point group. 
\end{abstract}

\maketitle

Quantum spin liquids are conjectured states of matter that have no long-range magnetic order and thus cannot be distinguished by their physical symmetries. The low-energy physics of spin liquids are often described in terms of partons---matter particles with fractional quantum numbers---interacting with emergent gauge fields \cite{PhysRevB.37.3774, PhysRevB.38.316, IJMPB.5.219}. Wen \cite{PhysRevB.65.165113} proposed to classify spin liquids on the basis of \emph{projective} symmetry, a combination of physical and gauge symmetries. Unfortunately, solvable models of spin liquids in more than one spatial dimension are hard to find. For this reason, partons and gauge fields in spin models have been typically introduced by fiat: spin variables are expressed in terms of Abrikosov fermions or Schwinger bosons and the resulting Hamiltonian, quartic in parton fields, is treated at the mean-field level. Although this approach can be justified in some limits, e.g., by taking the number of parton flavors $N \to \infty$ \cite{PhysRevB.37.3774, PhysRevB.38.316, IJMPB.5.219}, its applicability to physical spin models is debatable. 

The association of projective symmetry with \emph{ad hoc} fractionalization schemes \cite{PhysRevB.65.165113, PhysRevB.74.174423, PhysRevB.83.224413, PhysRevB.84.094419, PhysRevB.87.104406} is unfortunate. It is therefore desirable to find clean applications of projective symmetry to exactly solvable models of spin liquids. To that end we show that the properties of partons in Kitaev's honeycomb spin model \cite{AnnPhys.321.2} are best characterized in the language of projective symmetry. 

\begin{figure}
\includegraphics[width=0.95\columnwidth]{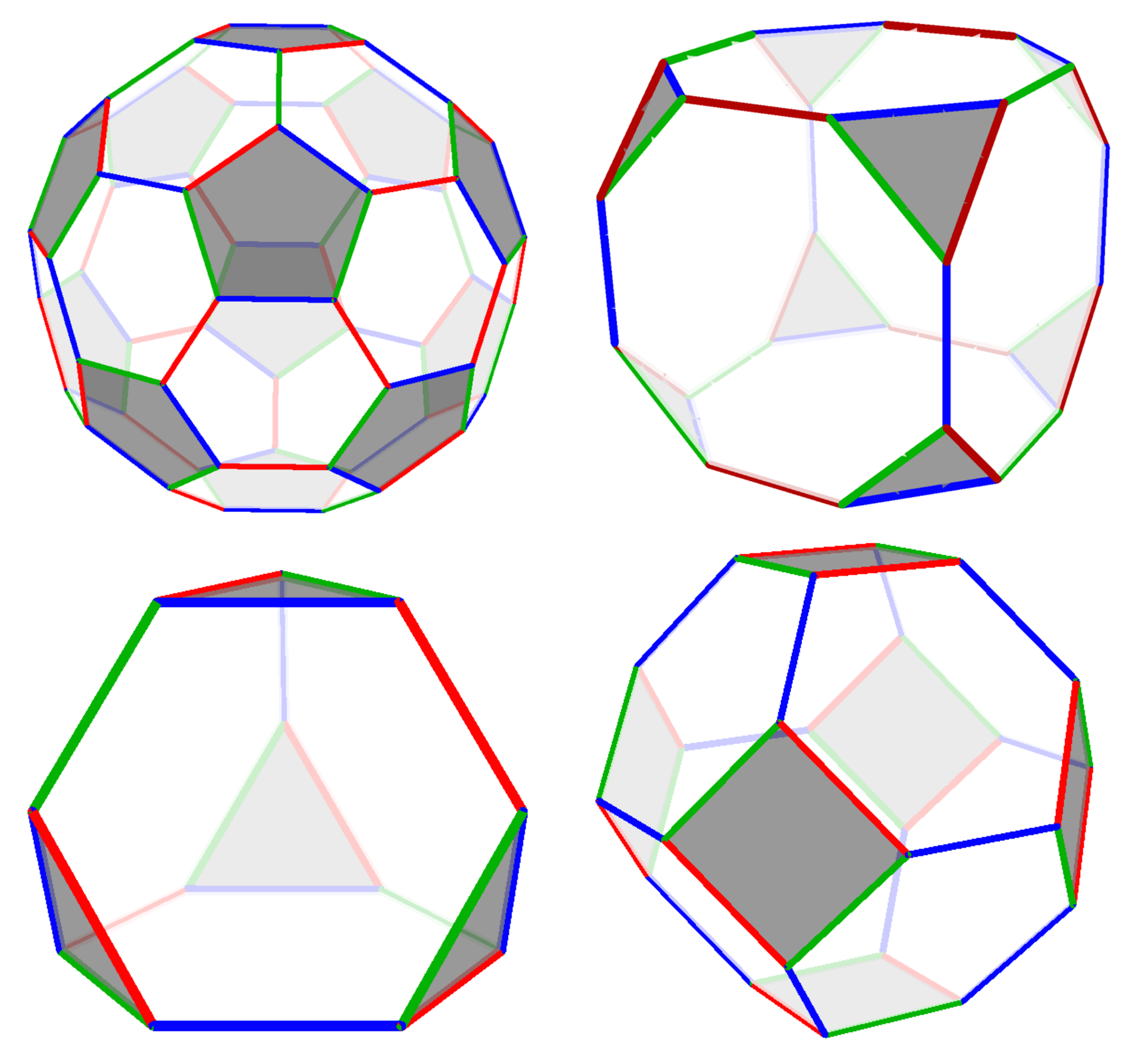}
\caption{Truncated tetrahedron, octahedron, cube, and icosahedron. Red, green, and blue edges have spin flavors $x$, $y$, and $z$, respectively. Shaded faces contain nonzero magnetic flux in the ground state.}
\label{fig:clusters}
\end{figure}

\emph{Summary of main results.} We study Kitaev's honeycomb spin model in a spherical lattice geometry realized by Archimedean solids
, Fig.~\ref{fig:clusters}. The model is solvable by a fermionization procedure yielding a Hamiltonian quadratic in Majorana fermions $c_n$. Naturally, the spectrum of a highly symmetric cluster is degenerate. 
However, the multiplicities do not match the dimensions of the point-group irreps. For example, the group of tetrahedron $T$ has irreps $\bm 1$, $\bm{1}'$, $\bm{1}''$, and $\bm{3}$, labeled by their dimensions (we follow the notation of \textcite{JPhysA.45.233001}). Unexpectedly, Majorana modes on a truncated tetrahedron form doublets (Table~\ref{table:1}), which the point-group symmetries fail to explain.

The resolution of this paradox is tied to the presence of a gauge field felt by Majorana fermions. The net outward magnetic flux through plaquettes of the cluster is $\Phi = 4\pi g$, where $g$ can be interpreted as the charge of a magnetic monopole at the cluster's center. The orbital angular momentum of a parton with unit electric charge is incremented by the angular momentum of the electromagnetic field $g$ \cite{Shnir.2005}. Because $g$ is half-integer in the ground states of our clusters, the net angular momentum is converted from integer to half-integer. To accommodate states with half-integer angular momenta, we must enlarge the point group $T \subset SO(3)$ to its double cover $\tilde{T} \subset SU(2)$ and use the irreps for which a rotation through $2\pi$ yields a factor of $-1$ \cite{LandauIII}. The double group $\tilde{T}$ has three such irreps: $\bm{2}$, $\bm{2}'$, and $\bm{2}''$. Hence the parton doublets. Similar scenarios apply to other spherical clusters: the projective symmetry group $\mathcal G$ of the ground state turns out to be the double cover $\tilde{G} \subset SU(2)$ of the corresponding point symmetry group $G \subset SO(3)$.  

\begin{table}[bt]
\begin{tabular}{|l|c|c|c|c|}
\hline
Solid & Multiplicities & $\Phi$ & $g$ & PSG\\
\hline
Truncated tetrahedron 
	& 2, 2, 2
	& $2\pi$ & $1/2$ & $\tilde{T}$ 
	\\ 
\hline
Truncated octahedron 
	& 4, 4, 4
	& $6 \pi$ & 3/2 & $\tilde{O}$ 
	\\
\hline
Truncated cube 
	& 4, 2, 4, 2
	& $2\pi$ & 1/2 &$\tilde{O}$ 
	\\
\hline
Truncated icosahedron 
	& 6, 2, 4, 6, 2, 6, 4
	& $6\pi$ & 3/2 & $\tilde{I}$ 
	\\
\hline
\end{tabular}
\caption{Multiplicities of Majorana modes (in the order of increasing energy $\epsilon>0$), net magnetic flux $\Phi$, monopole charge $g$, and projective symmetry group for Kitaev's spin model on some Archimedean solids. 
}
\label{table:1}
\end{table}

\emph{Landau levels on a sphere.} Before turning our attention to Kitaev's lattice model, we illustrate the relevant concepts in a related continuum problem: Landau levels of a massive particle on a sphere \cite{PhysRevLett.51.605}. It is convenient to treat it as a rigid rotor---a particle pivoted on a massless rod of length $r$---with mutually orthogonal axes $\hat{\bm \xi}, \hat{\bm \eta}, \hat{\bm \zeta}$ affixed to it; in particular, $\hat{\bm \zeta} = \mathbf r/r$ points along the rod. Note that body components of orbital angular momentum $L_\xi$, $L_\eta$, and $L_\zeta$ commute with the global components $L_x$, $L_y$, and $L_z$, so we may use as basis vectors simultaneous eigenstates of $L^2$, $L_z$, and $L_\zeta$ \cite{LandauIII}. 
The Hamiltonian is $H = L_\xi^2/2I_\xi + L_\eta^2/2I_\eta + L_\zeta^2/2I_\zeta$, where $I_\xi = I_\eta = mr^2$ and $I_\zeta = 0$. The vanishing of $I_\zeta$ requires setting $L_\zeta = 0$ in order to keep the energy finite, so
$H = (L_\xi^2 + L_\eta^2)/2mr^2 = L^2/2mr^2$.
In the presence of a magnetic field, the Hamiltonian is modified by replacing $\mathbf L = \mathbf r \times \mathbf p$ with 
$\bm \Lambda = \mathbf r \times (\mathbf p - \mathbf A) = \mathbf L - \mathbf r \times \mathbf A$, 
where $\mathbf A$ is the vector potential. 

Although magnetic field in this problem, $\mathbf B(\mathbf r) = g \mathbf r/r^3$, is spherically symmetric, the vector potential $\mathbf A(\mathbf r)$ is not. We can undo the change induced in $\mathbf A(\mathbf r)$ by a rotation if we follow it up with a gauge transformation \cite{SupMat}. The combined operation---a gauged rotation---leaves the vector potential, and thus the Hamiltonian, invariant. The generator of gauged rotations, 
\begin{equation}
\mathbf J = \mathbf L -  \mathbf r \times \mathbf A - g \mathbf r/r = \bm \Lambda - g \hat{\bm \zeta},
\end{equation}
satisfies the standard algebra of angular momentum \cite{PhysRevLett.51.605}. Its body-axis component $J_\zeta = L_\zeta - g = -g$. This constraint restricts $g$ to integer and half-integer values and the length of the gauged angular momentum to $j = |g|, |g|+1, |g|+2, \ldots$ The Hamiltonian, expressed in terms of $\mathbf J$, reads 
\begin{equation}
H = (\Lambda_\xi^2 + \Lambda_\eta^2)/2mr^2 = (\mathbf J^2 - g^2)/2mr^2.
\end{equation}

\emph{Kitaev's lattice model.} The Hamiltonian of Kitaev's spin model is \cite{AnnPhys.321.2}
\begin{equation}
H = - \sum_{\langle mn \rangle} J_{mn} S_m^{\alpha(mn)} S_n^{\alpha(mn)},
\label{eq:H-spin}
\end{equation}
where $\langle mn \rangle$ denotes a pair of nearest-neighbor sites $m$ and $n$ with a coupling constant $J_{mn}$. The spin component, or flavor, $\alpha(mn) = x$, $y$, or $z$ depends on the link $\langle mn \rangle$. With spins $\mathbf S_n$ represented in terms of four Majorana fermions $b_n^\alpha$, and $c_n$, $S_n^\alpha = i b_n^\alpha c_n$, the Hamiltonian becomes quadratic in $c$ fermions:
\begin{equation}
H =  - \sum_m \sum_n t_{mn} c_m c_n/4.
\label{eq:H-Majorana}
\end{equation}
Two $b$ fermions sharing a link combine to form a $Z_2$ gauge variable $u_{mn} = i b_{m}^{\alpha(mn)} b_{n}^{\alpha(mn)} = - u_{nm}$. Link variables $u$ commute with each other and with the Hamiltonian (\ref{eq:H-Majorana}) and can therefore be treated as numbers $u_{mn} = \pm 1$. The hopping matrix of $c$ Majorana fermions $t_{mn} = - 2 i J_{mn} u_{mn}$ is pure imaginary, antisymmetric, and thus Hermitian. 

We work with Archimedean solids obtained from Platonic solids by truncation, Fig.~\ref{fig:clusters}. Without loss of generality, we use ferromagnetic coupling constants, $J_1 > 0$ on edges inherited from Platonic solids and $J_2 > J_1 > 0$ on the edges resulting from truncation.

The product of link variables around a loop gives the $Z_2$ magnetic flux $W = (-i u_{12}) (- i u_{23}) \ldots (-i u_{L1})$. The allowed values of the flux depend on the perimeter $L$ of the loop: $W = \pm 1$ for even $L$ and $\pm i$ for odd $L$ \cite{AnnPhys.321.2, arXiv1406.6407}. Distinct physical states of the spin model can be fully specified by the values of $Z_2$ fluxes on all plaquettes and by the state of the $c$ Majorana fermions in this static magnetic background. Different  gauge representations $\{u\}$ of the same flux pattern $\{W\}$ are related by a gauge transformation
\begin{equation}
u_{mn}' = \Lambda_m u_{mn} \Lambda_n, 
\quad 
c_n' = \Lambda_n c_n,
\quad
\Lambda_n = \pm 1.
\label{eq:gauge-transform}
\end{equation}
The physics of the $Z_2$ gauge field in Kitaev's model has been explored in Refs.~\onlinecite{PhysRevB.84.115146, PhysRevLett.112.207203, arXiv1309.1171, PhysRevLett.113.197205}.

The Hamiltonian of the Majorana operators (\ref{eq:H-Majorana}) can be reduced to a diagonal form,
\begin{equation}
H = \sum_k \epsilon_k ( \gamma_k^\dagger \gamma_k - \gamma_k \gamma_k^\dagger )/2,
\label{eq:H-complex}
\end{equation}
where $\gamma_k = \frac{1}{2}\sum_n \psi_n^{(k)} c_n$ and $\gamma_k^\dagger$ are annihilation and creation operators of (complex) fermion eigenmodes and $\epsilon_k \geq 0$ are their excitation energies. The one-particle wavefunctions $\psi_n^{(k)}$ and the eigenvalues $\epsilon_k$ can be found by solving the one-particle Schr{\"o}dinger equation $-\sum_n t_{mn} \psi_n = \epsilon \psi_m$ with a pure imaginary hopping amplitude $t_{mn} = - 2i J_{mn} u_{mn}$ \cite{AnnPhys.321.2}. Eigenvalues of $t_{mn}$ come in pairs $\pm \epsilon$: if wavefunction $\psi_n$ has the eigenvalue $+\epsilon$ then its complex conjugate $\psi_n^*$ has the eigenvalue $- \epsilon$. Positive eigenvalues are the excitation energies of the Majorana eigenmodes in Eq.~(\ref{eq:H-complex}). The $Z_2$ flux $W = e^{i \Phi}$ translates into a $U(1)$ flux $\Phi$ experienced by these fermions. The allowed values of $\Phi$ depend on the loop length $L$: 0 and $\pi$ for $L$ even, $\pm \pi/2$ for $L$ odd.

The flux pattern in the ground state can be found from the following heuristic rules \cite{arXiv1406.6407}. A loop with an odd perimeter $L$ is indifferent to the value of its flux $\Phi = \pm\pi/2$. For even $L$, there is a preferred value:  $\Phi = 0$ if $L = 2 \mathop{\mathrm{mod}} 4$ and $\Phi = \pi$ if $L = 0 \mathop{\mathrm{mod}} 4$. For example, the ground state of the honeycomb model has zero flux on all hexagons \cite{AnnPhys.321.2}.

\begin{figure}
\includegraphics[width=0.95\columnwidth]{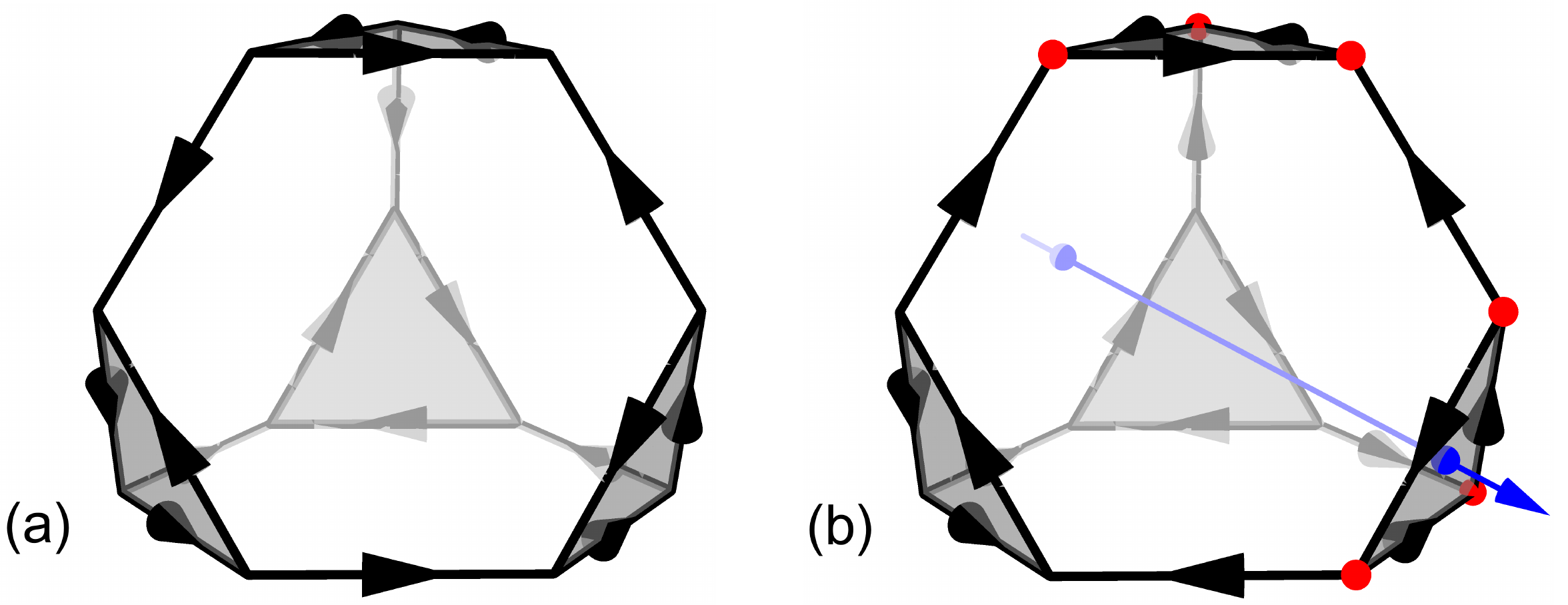}
\caption{(a) A ground state of Kitaev's model on a truncated tetrahedron. Arrows show directions for which the phase of the hopping amplitudes $\mathop{\mathrm{arg}}{t} = - \pi/2$. (b) The same state rotated through $\frac{2\pi}{3}$ about the indicated threefold symmetry axis. The hopping amplitudes can be restored by a $Z_2$ gauge transformation on vertices labeled with dots.}
\label{fig:truncated-tetrahedron}
\end{figure}

\emph{Projective symmetry group.}
We next construct the projective symmetry group (PSG) for the truncated tetrahedron. In the ground state, its hexagons have no flux, whereas all triangles have the same flux $\Phi = +\pi/2$ or $-\pi/2$. The net flux $\Phi = \pm 2\pi$ means a half-integer monopole charge $g = \Phi/(4\pi) = \pm 1/2$.  A gauge configuration $\{u\}$ for one of the two ground states is shown in Fig.~\ref{fig:truncated-tetrahedron}(a). The presence of a gauge field endows edges with a sense of direction and thereby reduces the symmetry.

Consider rotation $R(\frac{2\pi}{3},\hat{\mathbf n})$ about threefold axis $\hat{\mathbf n}$ [Fig.~\ref{fig:truncated-tetrahedron}(b)] that reverses the sign for some of the gauge variables $u_{mn}$ and the corresponding hopping amplitudes $t_{mn}$. As in Haldane's problem, the flux pattern remains unchanged and we may restore the original $u$ and $t$ by a gauge transformation (\ref{eq:gauge-transform}). One such transformation---let us call it $\Lambda(\frac{2\pi}{3},\hat{\mathbf n})$---has $\Lambda_n = -1$ on sites marked with red dots in Fig.~\ref{fig:truncated-tetrahedron}(b) and $+1$ on the remaining sites. The combined operation of gauged rotation, 
\begin{equation}
\textstyle
\mathcal R(\frac{2\pi}{3}, \hat{\mathbf n}) 
	= \Lambda(\frac{2\pi}{3}, \hat{\mathbf n}) \, R(\frac{2\pi}{3}, \hat{\mathbf n}),
\label{eq:gauged-rotation-120}
\end{equation}
leaves the hopping matrix invariant \cite{SupMat}. The complementary gauge transformation, $\Lambda'_n = - \Lambda_n$, also restores the gauge configuration. This is a general result: every point-group symmetry $R$ generates two gauged symmetries: $\Lambda R$ and $-\Lambda R$.

The $\frac{2\pi}{3}$ gauged rotation (\ref{eq:gauged-rotation-120}) has a peculiar property: applying it three times yields not the identity but rather multiplication by $-1$ \cite{SupMat}. If we identify this operation with a $2\pi$ gauged rotation then we find a result reminiscent of half-integer spin, $\mathcal R(2\pi, \hat{\mathbf n}) \equiv \mathcal R^3(\frac{2\pi}{3}, \hat{\mathbf n}) = -1$. Alternatively, we may define the gauged $2\pi$ rotation as a combination of the ordinary rotation $R(2\pi, \hat{\mathbf n}) = 1$ with the global gauge transformation $\Lambda(2\pi, \hat{\mathbf n}) = -1$: $\mathcal R(2\pi, \hat{\mathbf n}) \equiv \Lambda(2\pi, \hat{\mathbf n})R(2\pi, \hat{\mathbf n}) = -1$. Then the gauged symmetries satisfy the composition rule for rotations, $\mathcal R^3(\frac{2\pi}{3}, \hat{\mathbf n}) = \mathcal R(2\pi, \hat{\mathbf n})$.

The PSG for the ground-state flux sector is obtained as follows. We first construct two gauged rotations $\mathcal R(\frac{2\pi}{3}, \hat{\mathbf n}_1)$ and $\mathcal R(\frac{2\pi}{3}, \hat{\mathbf n}_2)$ about two different threefold axes $\hat{\mathbf n}_1$ and $\hat{\mathbf n}_2$ from ordinary rotations as described above. We then use the multiplication table of $SU(2)$ rotations (more precisely, of its subgroup $\tilde{T}$) to generate new elements and label them accordingly, e.g., $\mathcal R(\frac{2\pi}{3}, \hat{\mathbf n}_2) \mathcal R^{-1}(\frac{2\pi}{3}, \hat{\mathbf n}_1) = \mathcal R(\pi, \hat{\mathbf n}_3)$, where $\hat{\mathbf n}_3$ is a twofold axis.  We check that each new element $\mathcal R(\phi,\mathbf n)$ is indeed a gauged rotation, i.e., a composition of the corresponding ordinary rotation $R(\phi,\mathbf n) \in T$ and of a $Z_2$ gauge transformation $\Lambda(\phi,\mathbf n)$ defined in Eq.~(\ref{eq:gauge-transform}). Lastly we check that the multiplication tables of the newly constructed group and of $\tilde{T}$ are the same. This program establishes that the PSG of the ground-state flux sectors is indeed $\tilde{T}$, the double cover of the point group $T$. Similar results are obtained with the other Archimedean solids (Table~\ref{table:1}). 

\emph{Parton multiplets.} The number of vertices in a truncated Platonic solid equals the order of the corresponding point group $G \subset SO(3)$. States of a fermion living on the vertices transform under the regular representation of the group $G$. These states can be uniquely labeled by group elements as follows. Assign the identity element $e$ to an arbitrary vertex; the rest of the vertices are labeled by the group element $R \in G$ that takes vertex $e$ into them. (We now regard symmetries as active transformations.) A symmetry $R_2 \in G$ acts on state $|R_1\rangle$ as left multiplication: 
\begin{equation}
R_2 |R_1\rangle = |R_2 R_1\rangle. 
\label{eq:action-G}
\end{equation}
For the truncated tetrahedron and its group $G = T$, the regular representation is decomposed into irreps as follows: $\bm{12} = \bm 1 + \bm{1}' + \bm{1}'' + 3 \times \bm 3$. 

The same applies to the double cover $\tilde{G}$ of group $G$, except that each fermion state is now represented by group elements $\mathcal R \in \tilde{G}$ twice, as $\pm|\mathcal R\rangle$. The one-fermion states are decomposed into only those irreps of $\tilde{G}$ for which a $2\pi$ rotation equals multiplication by $-1$. For the double tetrahedral group $\tilde{T}$, $\bm{12} = 2 \times \bm{2} + 2 \times \bm{2}' + 2 \times \bm{2}''$. Thus we expect six doublets for a complex fermion on a truncated tetrahedron. For a Majorana fermion, states obtained by complex conjugation are identified and we obtain three doublets, as is indeed the case (Table~\ref{table:1}). For the double octahedral group $\tilde{O}$, $\bm{24} = 2 \times \tilde{\bm{2}} + 2 \times \tilde{\bm{2}}' + 4 \times \tilde{\bm{4}}$. Majorana fermions on a truncated cube indeed come in multiplets of 2, 2, 4, and 4.  On a truncated octahedron, the doublets are ``accidentally'' degenerate (Table~\ref{table:1}). For the buckyball (the icosahedral group $I$), $\bm{60} = 2 \times \tilde{\bm{2}} + 2 \times \tilde{\bm{2}}' + 4 \times \tilde{\bm{4}} + 6 \times \tilde{\bm{6}}$; we expect Majorana multiplets with dimensions 2, 2, 4, 4, 6, 6, 6, in agreement with direct diagonalization (Table~\ref{table:1}).

\begin{figure}
\includegraphics[width=0.95\columnwidth]{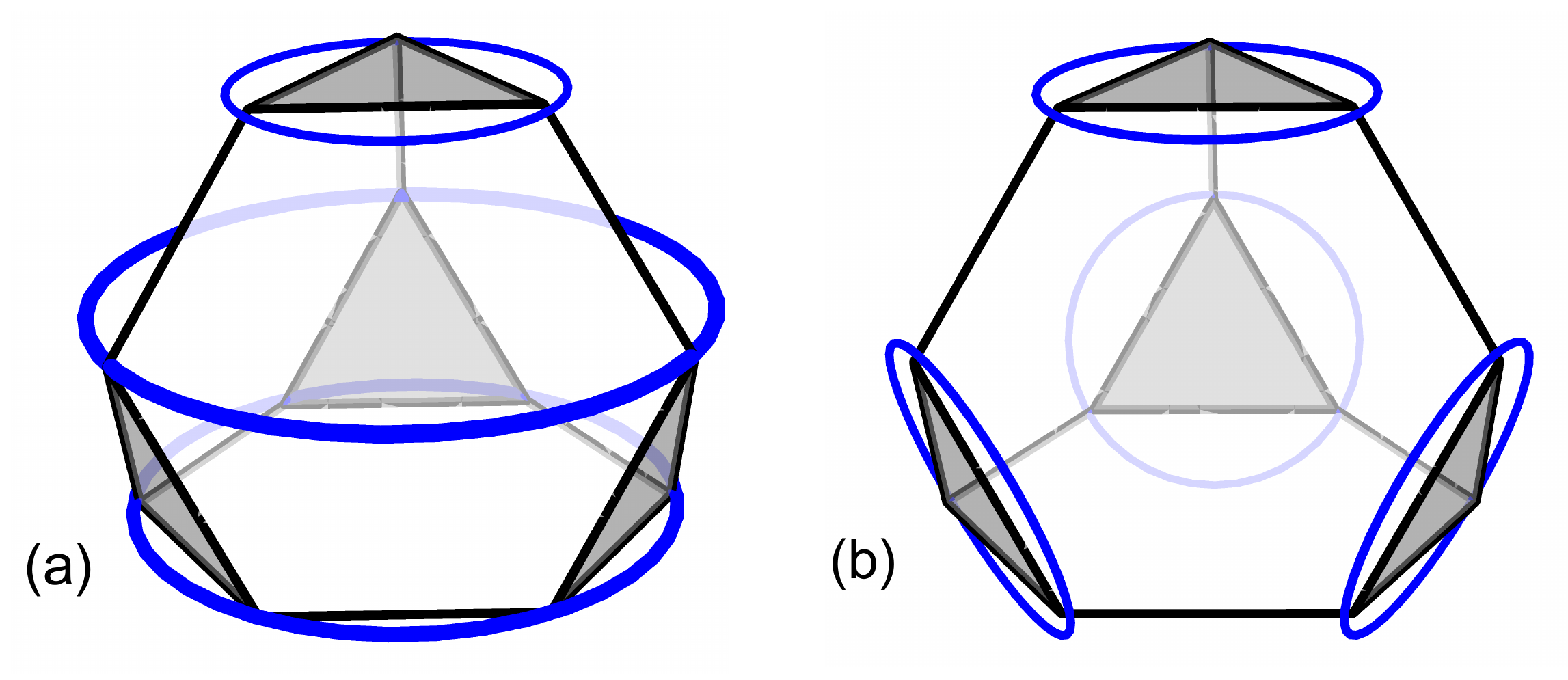}
\caption{Unprimed (a) and primed (b) rotations about a threefold axis. }
\label{fig:unprimed-primed}
\end{figure}

\emph{Parton spectrum.}
Symmetries $R \in G$ (or $\mathcal R \in \tilde{G}$) used so far represent rotations about axes fixed in space. It is convenient to introduce a second set of ``primed'' operations $R'$ (or their gauged versions $\mathcal R'$) representing rotations about axes that themselves rotate: 
\begin{equation}
R_2' |R_1\rangle = |R_1 R_2\rangle = R_1 R_2 R_1^{-1}|R_1\rangle. 
\label{eq:action-G-prime}
\end{equation}
If $R_2$ is a rotation about the axis nearest to vertex $e$ then $R_1 R_2 R_1^{-1}$ is an equivalent rotation about the axis nearest to vertex $R_1$, Fig.~\ref{fig:unprimed-primed}. 
The primed operations are direct analogs of rotations about axes attached to a rigid body, which also follow right multiplication \cite{Sakurai}.
The groups of primed and unprimed rotations are isomorphic: the multiplication table for $R'$ is the same as that of $R^{-1}$. Primed and unprimed rotations commute: $R_3 R_2'|R_1\rangle = |R_3 R_1 R_2\rangle = R_2' R_3|R_1\rangle$. 

As the hopping matrix $t$ commutes with unprimed rotations, we may guess that it can be expressed in terms of primed rotations. Indeed, for the truncated tetrahedron, it is a superposition of gauged rotations through $+\pi$ about the nearest twofold axis $\hat{\mathbf n}_1$ and through $+\frac{2\pi}{3}$ and $-\frac{2\pi}{3}$ about the nearest threefold axis $\hat{\mathbf n}_2$: 
\begin{equation}
\textstyle 
t = - 2i [J_1 \mathcal R'(\pi, \hat{\mathbf n}_1)
	 - J_2\mathcal R'(\frac{2\pi}{3}, \hat{\mathbf n}_2) 
	+ J_2 \mathcal R'(- \frac{2\pi}{3}, \hat{\mathbf n}_2)].
\label{eq:t-R}
\end{equation}
Because primed rotations form group $\tilde{T}$, we may use its irreps to block-diagonalize the hopping matrix. The block that corresponds to irrep $\lambda$ is obtained by replacing $\mathcal R'(\phi, \hat{\mathbf n})$ in Eq.~(\ref{eq:t-R}) with the irrep matrix $\mathcal D^{(\lambda)}(-\phi, \hat{\mathbf n})$. Matrices for irrep $\bm{2}$ of $\tilde{T}$ coincide with matrices of finite rotation of the fundamental (spin-$\frac{1}{2}$) irrep of $SU(2)$: $\mathcal D^{(\bm{2})}(-\phi, \hat{\mathbf n}) = e^{i (\bm{\sigma} \cdot \hat{\mathbf n})\phi/2}$, where $\bm{\sigma} = (\sigma_x, \sigma_y, \sigma_z)$ are the Pauli matrices. Taking the axes to be $\hat{\mathbf n}_1 = (0,0,1)$ and $\hat{\mathbf n}_2 = (1,1,1)/\sqrt{3}$, we obtain a $2 \times 2$ block $t^{(\bm{2})} = - 2 J_1 \sigma_z + 2 J_2(\sigma_x + \sigma_y + \sigma_z)$, whose positive eigenvalue $\epsilon = 2\sqrt{J_1^2 - 2 J_1 J_2 + 3J_2^2}$ matches the energy of one of the Majorana doublets obtained by direct diagonalization of the hopping matrix. Irreps $\bm{2}'$ and $\bm{2}''$ of $\tilde{T}$ cannot be expressed in terms of $SU(2)$ rotation matrices. However, their direct sum $\bm{2}'+\bm{2}''$ coincides with the 4-dimensional (spin-$\frac{3}{2}$) irrep of $SU(2)$, so we may again use $SU(2)$ rotation matrices 
to obtain a $4 \times 4$ block. Parton energies are roots of the characteristic polynomial $P(\epsilon) = \epsilon^4 - (3 J_1^2 + 2 J_1 J_2 + 2 J_2^2) \epsilon^2 + 16 (J_1 + J_2)^2 J_2^2$. They reproduce the energies of the two remaining Majorana doublets. This diagonalization procedure also works correctly for the ground states of the other spherical clusters listed in Table~\ref{table:1}. 

We can gain an additional insight into the spectrum of Majorana fermions by making a direct connection to Haldane's continuum model discussed above. If the hopping matrix $t$ on a cluster were real and positive, the state with the lowest energy $\epsilon < 0$ would be $\psi_n = 1$, a lattice analog of the $s$ state, followed by analogs of states with angular momenta $\ell = 1, 2, \ldots$ with multiplicities $2 \ell + 1$ until the continuum approximation breaks down \cite{AIPConfProc.273.336}. In the presence of a magnetic flux $\Phi = 4\pi g$ through the cluster, the energy eigenstates on the sphere have angular momenta $j = |g|, |g|+1, |g|+2, \ldots$ in the order of increasing energy with multiplicities $2j+1$. The same can be expected for the eigenstates of the hopping matrix with energies $\epsilon < 0$. Because the positive eigenvalues of the Majorana hopping matrix mirror the negative ones, we expect that the parton multiplet with the highest energy $\epsilon > 0$ will have angular momentum $j = |g|$, followed by multiplets with $j = |g|+1, |g|+2, \ldots$ until the continuum approximation breaks down. Indeed, e.g., on the buckyball, $g = 3/2$, the highest-energy partons form a quartet ($j = 3/2$) and a sextet ($j = 5/2$), see Table~\ref{table:1}. The octet ($j = 7/2$) is split into a doublet and a sextet by  deviations from spherical symmetry due to the lattice. 

We have shown that parton excitations in Kitaev's honeycomb model on highly symmetric spherical clusters have half-integer orbital angular momenta due to a nontrivial gauge background resembling the field of a magnetic monopole with a half-integer charge. 
 
The structure of parton multiplets can be understood in the framework of projective symmetries, which combine physical and gauge transformations. For all spherical clusters we have examined, the projective symmetry group for the ground state is the double cover $\tilde{G}$ of the point group $G$. As far as we know, this is the first application of projective symmetries in a solvable model of a spin liquid. 

\emph{Acknowledgments.} The authors thank Y. Wan for useful discussions. They acknowledge financial support from Fondecyt under Grant No. 11121397, Conicyt under Grant No. 79112004, and the Simons Foundation (P.M.); the Max Planck Society and the Alexander von Humboldt Foundation (O.P.); and the US DOE Grant No. DE-FG02-08ER46544 (O.T.). The Aspen Center for Physics, where part of this work was done, is supported by the US NSF Grant No. PHY-1066293.

\bibliographystyle{apsrev}

\pagebreak
\widetext
\begin{center}
\textbf{\large Supplementary Material for\\ ``Projective symmetry of partons in Kitaev's honeycomb model"}
\end{center}
\setcounter{equation}{0}
\setcounter{figure}{0}
\setcounter{table}{0}
\setcounter{page}{1}
\makeatletter
\renewcommand{\theequation}{S\arabic{equation}}
\renewcommand{\thefigure}{S\arabic{figure}}
\renewcommand{\bibnumfmt}[1]{[S#1]}
\renewcommand{\citenumfont}[1]{S#1}

\section{PSG of the Regular Tetrahedron}

\subsection{Symmetries of a regular tetrahedron}

\begin{figure}[th]
\includegraphics[width=0.95\columnwidth]{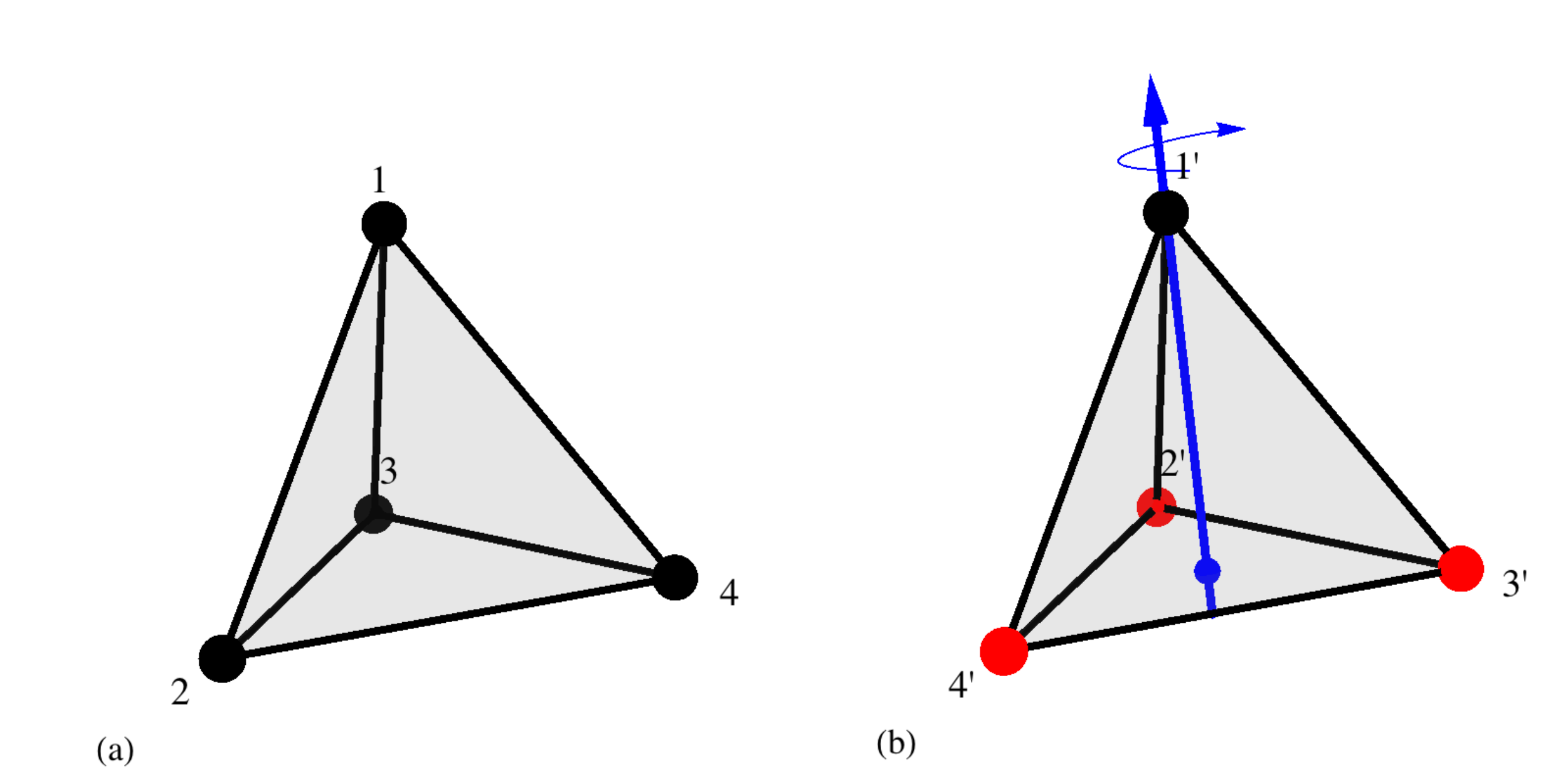}
\caption{a) A regular tetrahedron. b) A passive transformation: the reference frame (the labels) is rotated through $-2\pi/3$ about axis 1.}
\label{fig:S_tetrahedron0}
\end{figure}
Consider a complex fermion hopping between sites of a regular tetrahedron,
\[
H = - \sum_{m=1}^4\sum_{n \neq m} t_{mn} \chi_m^\dagger \chi_n. 
\]
A symmetry transformation maps a lattice into itself. Instead of active transformations such as a rotation of the lattice through angle $\phi$, we will use passive transformations, in which the reference frame rotates in the opposite direction, through angle $-\phi$. Fig.~\ref{fig:S_tetrahedron0} depicts a $+\frac{2\pi}{3}$ rotation about axis $\hat{\mathbf n}_1$ connecting the center of the tetrahedron and site 1; in the passive convention, the labels are rotated through angle $-\frac{2\pi}{3}$. Sites 1, 2, 3, and 4 are relabeled $1'$, $4'$, $2'$ and $3'$, respectively. We can express the transformation with a permutation matrix $R$ whose element $R_{mn} = 1$ if $m' = n$ and 0 otherwise: 
\begin{equation}
\textstyle
R(\frac{2\pi}{3}, \hat{\mathbf n}_1) = 
	\left(
		\begin{array}{cccc}
			1 & 0 & 0&0 \\
			0 & 0 & 0&1 \\
			0 & 1 & 0&0 \\
			0 & 0 & 1&0 
		\end{array}
	\right).
\end{equation}
This matrix is orthogonal, $R^T = R^{-1}$. Every row and every column contain only one nonzero entry. The new hopping amplitude $t'_{m'n'}$ is equal to the old hopping amplitude $t_{mn}$: 
\[
t'_{m'n'} = \sum_m \sum_n R_{m'm} R_{n'n} t_{mn} = (R \, t \, R^T)_{m'n'}.
\]
Since matrix $R$ is real, we may replace $R^T$ with $R^\dagger$:
\begin{equation}
t' = R \, t \, R^\dagger.
\end{equation}

Let us take all hopping matrix elements to be equal: 
\begin{equation}
t = \left(
		\begin{array}{cccc}
			0 & 1 & 1 & 1 \\
			1 & 0 & 1 & 1\\
			1 & 1 & 0 & 1\\
			1 & 1 & 1 & 0
		\end{array}
	\right).
\label{eq:S_t-triangle}
\end{equation}
The symmetry group of the hopping matrix includes four conjugacy classes: the identity $\{e\}$, four rotations $\{R(\frac{2\pi}{3}, \hat{\mathbf n})\}$ about axes $\hat{\mathbf n} = \hat{\mathbf n}_1$ through $ \hat{\mathbf n}_4$ connecting the center of the tetrahedron with the respective vertices, their inverses $\{R(-\frac{2\pi}{3}, \hat{\mathbf n})\}$, and three rotations $\{R(\pi, \hat{\bm \alpha})\}$ about axes $\hat{\bm \alpha} = x$, $y$, and $z$ connecting the centers of opposite edges:
\begin{eqnarray}
&
\textstyle
R(\frac{2\pi}{3}, \hat{\mathbf n}_1) = 
	\left(
		\begin{array}{cccc}
			1 & 0 & 0&0 \\
			0 & 0 & 0&1 \\
			0 & 1 & 0&0 \\
			0 & 0 &1&0 
		\end{array}
	\right),
\ 
R(\frac{2\pi}{3}, \hat{\mathbf n}_2) = 
	\left(
		\begin{array}{cccc}
			0 & 0 & 1&0 \\
			0 & 1 & 0&0 \\
			0 & 0 & 0&1 \\
			1 & 0 & 0&0 
		\end{array}
	\right),
\ 
R(\frac{2\pi}{3}, \hat{\mathbf n}_3)= \left(
		\begin{array}{cccc}
			0 & 0 & 0&1 \\
			1 & 0 & 0&0 \\
			0 & 0 & 1&0 \\
			0 & 1 & 0&0 
		\end{array}
	\right),
\ 
R(\frac{2\pi}{3}, \hat{\mathbf n}_4)= \left(
		\begin{array}{cccc}
			0 & 1 & 0&0 \\
			0 & 0 & 1&0 \\
			1 & 0 & 0&0 \\
			0 & 0 & 0&1 
		\end{array}
	\right),
&
\nonumber\\
&
R(\pi, \hat{\mathbf x})= \left(
		\begin{array}{cccc}
			0 & 0 & 0&1 \\
			0 & 0 & 1&0 \\
			0 & 1 & 0&0 \\
			1 & 0 & 0&0 
		\end{array}
	\right),
\quad
R(\pi, \hat{\mathbf y})= \left(
		\begin{array}{cccc}
			0 & 0 & 1&0 \\
			0 & 0 & 0&1 \\
			1 & 0 & 0&0 \\
			0 & 1 & 0&0 
		\end{array}
	\right),
\quad
R(\pi, \hat{\mathbf z})= \left(
		\begin{array}{cccc}
			0 & 1 & 0&0 \\
			1 & 0 & 0&0 \\
			0 & 0 & 0&1 \\
			0 & 0 & 1&0 
		\end{array}
	\right).
&
\end{eqnarray}
Its characters are listed in Table~\ref{table:S_tetrahedron0-group}. The group has irreps $\bm 1$, $\bm 1'$, $\bm 1''$, and $\bm 3$, labeled by their dimensions. The fermion annihilation operators $\{\chi_n\}$ transform in terms of each other under these symmetries. 

\begin{table}
\begin{tabular}{{|l||c|c|c|c|c|}}
\hline
irrep & $E$ & $4C_3$ & $4(C_3)^2$ & $3C_2$\\
\hline\hline
1 & 1 & 1 & 1  & 1 \\
\hline
$1'$ & 1 & $\omega$ & $\omega^*$ & 1 \\
\hline
$1''$ &1& $\omega^*$ & $\omega$ & 1 \\
\hline
3 & 3 & 0 & 0 & -1\\
\hline
\end{tabular}
\caption{Characters of  irreps of the tetrahedral group $T$. $\omega = e^{2\pi i/3}$.}
\label{table:S_tetrahedron0-group}
\end{table}

\subsection{Gauge transformations}

\begin{figure}[bh!]
\includegraphics[width=0.95\columnwidth]{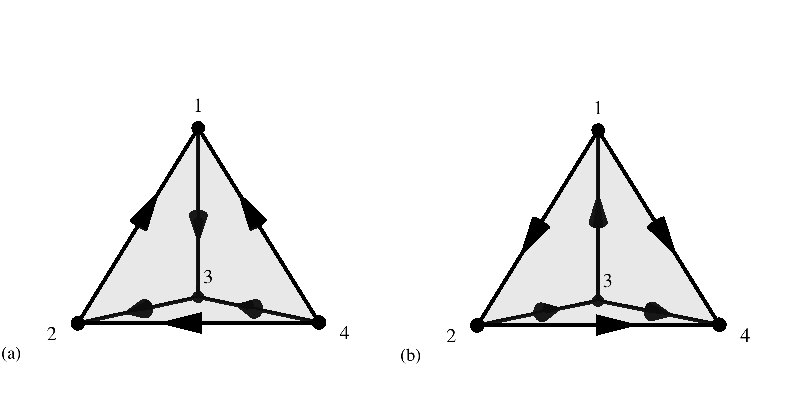}
\caption{Kitaev's model on a tetrahedron. Arrows indicate edge directions along which the hopping amplitude is $i$. The outbound magnetic flux through every face of the tetrahedron is (a) $+\pi/2$ and (b) $-\pi/2$.}
\label{fig:S_tetrahedron}
\end{figure}

\begin{figure}[bh!]
\includegraphics[width=0.45\columnwidth]{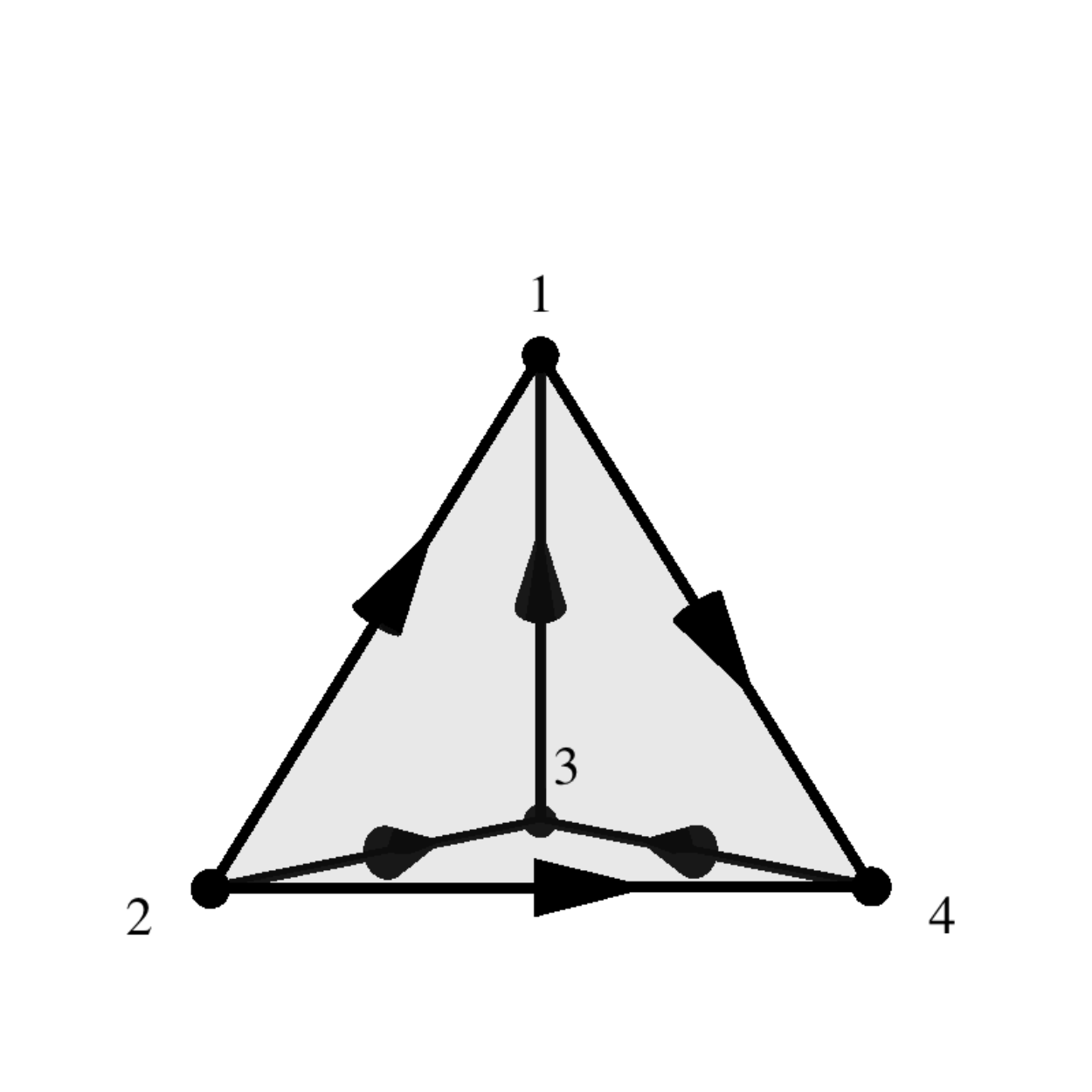}
\caption{Kitaev's model on tetrahedron of Fig.~\ref{fig:S_tetrahedron} (a) after a rotation in $2\pi/3$ clockwise around axis 1 has been performed.}
\label{fig:S_tetrahedron_rotated}
\end{figure}

Sometimes the Hamiltonian has a lower symmetry than the physical system itself. This situation often arises in the presence of gauge variables, which can change values while leaving the physical system invariant. Let us examine the case of the Kitaev model defined on a tetrahedron, i.e. the tight binding Hamiltonian for Majorana fermions,
\begin{equation}
H = - \sum_m \sum_n t_{mn}c_m c_n/4,
\label{eq:S_H-c}
\end{equation}
In order for the Hamiltonian to be Hermitian, the hopping amplitudes of Majorana fermions must be pure imaginary. Specifically, we consider a tetrahedron with hopping amplitudes of the same magnitude $\pm 2i$ as shown in Fig.~\ref{fig:S_tetrahedron}. The amplitudes are $2i$ in the directions shown by arrows. The hopping matrix of the tetrahedron in Fig.~\ref{fig:S_tetrahedron}(a) is
\begin{equation}
t = 
	2 \left(
		\begin{array}{cccc}
			 0 & -i & +i & -i \\
			+i &  0 & -i & -i \\
			-i & +i &  0 & -i \\
			+i & +i & +i &  0
		\end{array}
	\right)
\label{eq:S_t-tetrahedron}
\end{equation}
Going counterclockwise around a face of the tetrahedron, a fermion experiences a net flux $\Phi = -\pi/2$. In the absence of magnetic background, an arbitrary fermion wavefunction with amplitude on the four vertices would form a 4-dimensional representation of $T$ containing irreps $\bm 1$ and $\bm 3$. However diagonalization of the hopping matrix $t$ yields energy eigenvalues $+2\sqrt{3}$ and $-2\sqrt{3}$ both with multiplicity 2, rather than 1 and 3. Therefore the point group $T$ discussed in the previous section is not the symmetry group of the Hamiltonian Eq.~(\ref{eq:S_H-c}). However, the symmetry of the Hamiltonian can be restored if we follow the symmetry operations of $T$ with gauge transformations. For instance, rotation $R(\frac{2\pi}{3}, \hat{\mathbf n}_1)$ alters the signs of hopping on bonds 13, 14, 23 and 24 as shown in Fig.~\ref{fig:S_tetrahedron_rotated} . The change can be rectified by the gauge transformation $c_1 \mapsto - c_1$,  $c_2 \mapsto - c_2$, which we will name $ \Lambda(\frac{2\pi}{3}, \hat{\mathbf n}_1)$. This gauge transformation is expressed by a diagonal matrix: 
\[
\textstyle
 \Lambda(\frac{2\pi}{3}, \hat{\mathbf n}_1) = \left(
		\begin{array}{cccc}
			-1 & 0 & 0 &0\\
			0 & -1 & 0 &0\\
			0 & 0 & 1&0\\
			0 & 0 & 0&1
		\end{array}
	\right) 
	\equiv \mathop{\mathrm{diag}}{(-1,-1,1,1)}.
\]
It is easy to check that the combined transformation $ \Lambda R$ leaves the hopping matrix unchanged: 
\[
\textstyle
\Lambda(\frac{2\pi}{3}, \hat{\mathbf n}_1) 
R(\frac{2\pi}{3}, \hat{\mathbf n}_1) 
\, t \, 
R^\dagger (\frac{2\pi}{3}, \hat{\mathbf n}_1) 
\Lambda^\dagger(\frac{2\pi}{3}, \hat{\mathbf n}_1) 
= t.
\] 
The same correcting action can be achieved by another gauge transformation $-\Lambda(\frac{2\pi}{3}, \hat{\mathbf n}_1)$, which alters the signs of Majorana fermions $c_3$ and $c_4$, so $-\Lambda(\frac{2\pi}{3}, \hat{\mathbf n}_1) R(\frac{2\pi}{3}, \hat{\mathbf n}_1)$ is another symmetry of the tetrahedron. In fact, any combined symmetry+gauge operation now has a twin obtained by acting with the global gauge transformation $\Lambda(2\pi, \hat{\mathbf n}) = Q = -1$, which alters the signs of all fermion operators, $c_m \mapsto - c_m$. 

\subsection{Double cover of the tetrahedral group $\tilde{T}$}

Gauged rotation of the reference frame about the axis connecting the center and site 1,  
\begin{equation}
\textstyle
\mathcal{R}(\frac{2\pi}{3}, \hat{\mathbf n}_1)=\Lambda(\frac{2\pi}{3}, \hat{\mathbf n}_1)R(\frac{2\pi}{3}, \hat{\mathbf n}_1),
\end{equation}
leaves the Hamiltonian invariant. Transformations of this type form the projective symmetry group (PSG) of a given flux state. 

An important difference between regular rotations $R(\frac{2\pi}{3})$ and their gauged versions $\mathcal R(\frac{2\pi}{3})$ is that $R^3(\frac{2\pi}{3}) = R(2\pi) = 1$, whereas $\mathcal{R}^3(\frac{2\pi}{3})=-1$. This is precisely what one expects from a $2\pi$ rotation of a half-integer spin $\mathbf J$: $e^{- 2\pi i \, \mathbf J \cdot \hat{\mathbf n}} = -1$. We therefore expect that the PSG of the majorana fermions $c_n$ to be given by the generalization of the tetrahedral group $T$ to half-integer spins, similarly to the way that $SU(2)$, the double cover of the rotational group $SO(3)$,  includes irreps with even dimensions (half-integer spin lengths), in addition to irreps with odd dimensions (integer spin lengths). Indeed symmetries of the Kitaev model on a tetrahedron turn out to be described by the double cover of the tetrahedral group $\tilde{T}$.
\color{black}

Operation $\Lambda(\frac{2\pi}{3}, \hat{\mathbf n}_1)R(\frac{2\pi}{3}, \hat{\mathbf n}_1)$ can be identified as a $2\pi/3$ rotation of the double group $\mathcal{R}(\frac{2\pi}{3}, \hat{\mathbf n}_1)$. Its partner $-\Lambda(\frac{2\pi}{3}, \hat{\mathbf n}_1) R(\frac{2\pi}{3}, \hat{\mathbf n}_1)$ does not yield $-1$ when cubed and thus cannot be designated $\mathcal R(\frac{2\pi}{3}, \hat{\mathbf n}_1)$.   Next we proceed to find the rest of the PSG members. A clockwise rotation about axis connecting the center and the site 2 is 
\[
\textstyle
R(\frac{2\pi}{3}, \hat{\mathbf n}_2) = \left(
\begin{array}{cccc}
 0 & 0 & 1 & 0 \\
 0 & 1 & 0 & 0 \\
 0 & 0 & 0 & 1 \\
 1 & 0 & 0 & 0 
\end{array}
\right),
\]
it must be followed by either $\Lambda(\frac{2\pi}{3}, \hat{\mathbf n}_2)$ or $\Lambda(2\pi, \hat{\mathbf n})\Lambda(\frac{2\pi}{3}, \hat{\mathbf n}_2) = \Lambda(\frac{8\pi}{3}, \hat{\mathbf n}_2)$, the first gauge transformation changes the signs of Majoranas $c_2$ and $c_3$, while the second alters $c_1$ and $c_4$. The gauged $2\pi/3$ rotation is 
\[
\textstyle
\mathcal R(\frac{2\pi}{3}, \hat{\mathbf n}_2) 
	= \Lambda(\frac{2\pi}{3}, \hat{\mathbf n}_2)R(\frac{2\pi}{3}, \hat{\mathbf n}_2) 
	= \left(
		\begin{array}{cccc}
			0 & 0 & 1 & 0 \\
			0 & -1 & 0 & 0 \\
			0 & 0 & 0 & -1 \\
			1 & 0 & 0 & 0 
		\end{array}
	\right)
\]
There are a total of four gauged rotations 
\[
\textstyle
\mathcal R(\frac{2\pi}{3}, \hat{\mathbf n}_i) 
	= \Lambda(\frac{2\pi}{3}, \hat{\mathbf n}_i)R(\frac{2\pi}{3}, \hat{\mathbf n}_i)
\]
satisfying $\mathcal R^3(\frac{2\pi}{3}, \hat{\mathbf n}) = -1$. They form a conjugacy class as they can be transformed into one another, e.g.,
\[
\textstyle
\mathcal R(\frac{2\pi}{3}, \hat{\mathbf n}_1)\mathcal R(\frac{2\pi}{3}, \hat{\mathbf n}_2)\mathcal R^{-1}(\frac{2\pi}{3}, \hat{\mathbf n}_1)= \mathcal R(\frac{2\pi}{3}, \hat{\mathbf n}_3) .
\]
They can also be used as generators to build the rest of the group. For example, taking the $\mathcal R(\frac{2\pi}{3})$ elements to powers 2, 4, and 5 produces the conjugacy classes $\{\mathcal R^2(\frac{2\pi}{3}, \hat{\mathbf n}_i)\}$, $\{\mathcal R^4(\frac{2\pi}{3}, \hat{\mathbf n}_i)\} = \{\mathcal R^{-2}(\frac{2\pi}{3}, \hat{\mathbf n}_i)\}$, $\{\mathcal R^5(\frac{2\pi}{3}, \hat{\mathbf n}_i)\} = \{\mathcal R^{-1}(\frac{2\pi}{3}, \hat{\mathbf n}_i)\}$ of gauged rotations through angles $4\pi/3$, $8\pi/3 = - 4\pi/3$ (mod $4\pi$), and $10\pi/3 = - 2\pi/3$. The third power yields the $2\pi$ rotation $\mathcal R^3(\frac{2\pi}{3}, \hat{\mathbf n}_i) = Q = -1$, which is in a conjugacy class of its own. The sixth power (a $4\pi$ rotation) yields the identity element $e$. 

Twofold rotations of the double group $\mathcal R(\pi)$ can also be generated from threefold ones, e.g.,
\[
\textstyle
\mathcal R(\pi, \hat{\mathbf z}) = \mathcal R ^{-1}(\frac{2\pi}{3}, \hat{\mathbf n}_3)\mathcal R(\frac{2\pi}{3}, \hat{\mathbf n}_2).
\]
(This can be easily checked by multiplying finite rotation matrices $e^{- 2\pi i \, \mathbf J \cdot \hat{\mathbf n}}$ for $J=1/2$.) In this way we determine that 
\[
\mathcal R(\pi, \hat{\mathbf x}) = \Lambda(\pi, \hat{\mathbf x}) R(\pi, \hat{\mathbf x}),
\quad
\mathcal R(\pi, \hat{\mathbf y}) = \Lambda(\pi, \hat{\mathbf y}) R(\pi, \hat{\mathbf y}),
\quad
\mathcal R(\pi, \hat{\mathbf z}) = \Lambda(\pi, \hat{\mathbf z}) R(\pi, \hat{\mathbf z}).
\]
where 
\[
\Lambda(\pi, \hat{\mathbf x}) = \mathop{\mathrm{diag}}{(1,-1,1,-1)}, 
\quad
\Lambda(\pi, \hat{\mathbf y}) = \mathop{\mathrm{diag}}{(1,1,-1,-1)}, 
\quad
\Lambda(\pi, \hat{\mathbf z}) = \mathop{\mathrm{diag}}{(-1,1,1,-1)}. 
\]
These gauged symmetries and their inverses $\mathcal R^{-1}(\pi, \hat{\bm \alpha}) = \mathcal R^3(\pi, \hat{\bm \alpha})$ form the last conjugacy class $\{\mathcal R(\pm \pi, \hat{\bm \alpha})\}$. With the exception of threefold rotations $\mathcal R(\frac{2\pi}{3}, \hat{\mathbf n}_4) $ about the axis connecting the center and vertex 4, all regular symmetry operations change link variables and must be accompanied by a gauge transformation.  Therefore, the PSG contains 24 elements---twice as many as the point group $T$ of the tetrahedron. Each physical symmetry is augmented by a gauge transformation or by its complement. 

Fig.~\ref{fig:S_tetrahedron} (b), shows each triangular face having flux $-\pi/2$ measured counterclockwise from the outside of the tetrahedron. The net flux through the tetrahedron is $\Phi = 4 \times (- \pi/2) = - 2\pi$. Dividing the net flux by the full solid angle $4\pi$ gives the charge of the magnetic monopole at the center, $g = -1/2$. Since the charge is half-integer, the orbital angular momentum should also be half-integer \cite{S_PhysRevLett.51.605, S_PhysRevB.83.115129}.  We therefore expect that the PSG of the majorana fermions $c_n$ be described by the double cover of the tetrahedron group $\tilde{T}$. Double irreps of the tetrahedral group are 2-dimensional, so we expect twofold degeneracy of the fermion energies.

With 7 conjugacy classes, the double group $\tilde{T}$ has 7 irreps. 4 of these are the irreps of the regular tetrahedron group $T$ $\bm 1$, $\bm 1'$, $\bm 1''$, and $\bm 3$, labelled by their dimensions. The double irreps, odd under the global gauge $Q$, are $\bm 2$, $\bm 2'$, and $\bm 2''$. Their characters are listed in Table~\ref{table:S_double-T-group}. Irrep $\bm 2$ consists of finite rotation matrices for $J=1/2$. Finite rotation matrices for $J=3/2$ form a reducible representation that is the direct sum of irreps $\bm 2'$ and $\bm 2''$. 

\begin{table}
\begin{tabular}{|l||c|c|c|c|c|c|c|}
\hline
irrep & $e$ & $Q$ & $4\tilde{C}_3$ & $4\tilde{C}_3^2$ & $4\tilde{C}_3^{-2}$ & $4\tilde{C}_3^{-1}$ & $6\tilde{C}_2$\\
\hline\hline
$\bm 2$ & 2 & $-2$ & 1 & $-1$ & $-1$ & 1 & 0 \\
\hline
$\bm 2'$ & 2 & $-2$ & $\omega$ & $-\omega^*$ & $-\omega$ & $\omega^*$ & 0 \\
\hline
$\bm 2''$ & 2 & $-2$ & $\omega^*$ & $-\omega$ & $-\omega^*$ & $\omega$ & 0\\
\hline
\end{tabular}
\caption{Characters of double irreps of the double tetrahedral group $\tilde{T}$. $\omega = e^{2\pi i/3}$.}
\label{table:S_double-T-group}
\end{table}

Majorana fermions $c_n$ living on the sites of the tetrahedron form a reducible 4-dimensional representation of group $\tilde{T}$. It contains two 2-dimensional irreps. Its characters for $e$ and $Q$ are 4 and $-4$. Threefold gauged rotation $\mathcal R(\frac{2\pi}{3}, \hat{\mathbf n}_1) = \Lambda(\frac{2\pi}{3}, \hat{\mathbf n}_1)R(\frac{2\pi}{3}, \hat{\mathbf n}_1)$ leaves only one fermion, $c_1$, in place and reverses its sign. Thus the character of conjugacy class $\{\mathcal R(\frac{2\pi}{3}, \hat{\mathbf n}_i)\}$ is $-1$. From Table~\ref{table:S_double-T-group} one finds that these characters can only be obtained by taking the sum of irreps $\bm 2'$ and $\bm 2''$. So we expect two levels with degeneracy 2 each, as is indeed the case. 
 
\bibliographystyle{apsrev}

\end{document}